\documentclass[journal=tches,final,xcolor=table]{iacrtrans}

\usepackage{url}
\usepackage[table]{xcolor}
\usepackage{tabularx}
\newcolumntype{L}[1]{>{\raggedright\arraybackslash}p{#1}} 
\newcolumntype{C}[1]{>{\centering\arraybackslash}p{#1}} 
\newcolumntype{R}[1]{>{\raggedleft\arraybackslash}p{#1}} 
\usepackage{multirow}
\usepackage{adjustbox}
\usepackage{booktabs}
\usepackage{colortbl}

\usepackage{cite}
\usepackage{amsmath,amssymb,amsfonts}
\usepackage{algorithmic}
\usepackage{graphicx}
\usepackage{textcomp}
\usepackage[nolist]{acronym}

\begin{acronym}
\acro{BBRAM}{battery-backed RAM}
\acro{BRAM}{Block RAM}
\acro{CPA}{Correlation Power Analysis}
\acro{DSP}{Digital Signal Processor}
\acro{DoS}{Denial of Service}
\acro{DoD}{Department of Defense}
\acro{FF}{Flip-Flop}
\acro{FPGA}{Field Programmable Gate Array}
\acro{HD}{Hamming Distance}
\acro{HDL}{Hardware Description Language}
\acro{HW}{Hamming Weight}
\acro{IC}{Integrated Circuit}
\acro{ISA}{Instruction Set Architecture}
\acro{LUT}{Look-Up-Table} 
\acro{GCM}{Galois/Counter Mode}
\acro{GF}{Galois Field}
\acro{OTP}{One-Time-Programmable}
\acro{CTR}{Counter}
\acro{C-CTR}{Configuration Counter}

\acro{AES}{Advanced Encryption Standard}
\acro{AFL}{American Fuzzy Lop}
\acro{AI}{Artificial Intelligence}
\acro{ASIC}{Application-Specific Integrated Circuit}
\acro{BEL}{Basic Element}
\acro{BPI}{Byte Peripheral Interface}
\acro{CAN bus}{Controller Area Network}
\acro{CBC}{Cipher Block Chaining}
\acro{CLB}{Configurable Logic Block}
\acro{CLK}{Clock}
\acro{CMT}{Clock Management Tile}
\acro{CRC}{Cyclic Redundancy Check}
\acro{CVE}{Common Vulnerabilities and Exposures}
\acro{eFUSE}{Electronic Fuse}
\acro{FPGA}{Field Programmable Gate Array}
\acro{FSR}{Fabric Sub Region}
\acro{GSM}{Global System for Mobile Communications}
\acro{HAL}{Hardware Abstraction Layer}
\acro{HDL}{Hardware Description Language}
\acro{HLS}{High-Level Synthesis}
\acro{HMAC}{Hash-Based Message Authentication Code}
\acro{IC}{Integrated Circuit}
\acro{I/O}{Input/Output}
\acro{ICAP}{Internal Configuration Access Port}
\acro{IoT}{Internet of Things}
\acro{IV}{Initialization Vector}
\acro{JTAG}{Joint Test Action Group}
\acro{psbn}{post-start-bitstream-name}
\acro{RAM}{Random-Access Memory}
\acro{RSA}{Rivest–Shamir–Adleman}
\acro{RTL}{Register-Transfer Level}
\acro{SHA256}{Secure Hash Algorithm-256}
\acro{SHA-3}{Secure Hash Algorithm 3}
\acro{SLR}{Super Logic Region}
\acro{SPI}{Serial Peripheral Interface}
\acro{SSIT}{Stacked Silicon Interconnect Technology}
\acro{TCP}{Transmission Control Protocol}
\acro{UART}{Universal Asynchronous Receiver / Transmitter}
\acro{UDP}{User Datagram Protocol}

\end{acronym}

\definecolor{colKeys}{rgb}{0,0,1}
\definecolor{colIdentifier}{rgb}{0,0,0}
\definecolor{colComments}{rgb}{1,0,0}
\definecolor{colString}{rgb}{0,0.5,0}
\colorlet{gray_full}{black!30}
\colorlet{gray_mild}{black!15}

\usepackage{hyperref}

\usepackage{bytefield}

\usepackage{listings}
\usepackage{xspace}
\usepackage{comment}
\ifdefined\algorithmautorefname
\renewcommand{\algorithmautorefname}{Algorithm}
\fi 
\ifdefined\definitionautorefname
\renewcommand{\definitionautorefname}{Definition}
\fi

\usepackage{enumitem}   

\newcommand{\fwname}{\texttt{ConFuzz}\xspace}

\usepackage{tikz}

\newcommand{\circled}[2][]{
  \tikz[baseline=(char.base)]{
    \node[shape=circle,draw,inner sep=1pt,fill=black]
    (char) {\phantom{\ifblank{#1}{#2}{#1}}};
    \node[text=white] at (char.center) {\makebox[0pt][c]{\bfseries#2}};}\xspace}
\robustify{\circled}

\DeclareRobustCommand{\orcidicon}{%
	\includegraphics[width=4mm]{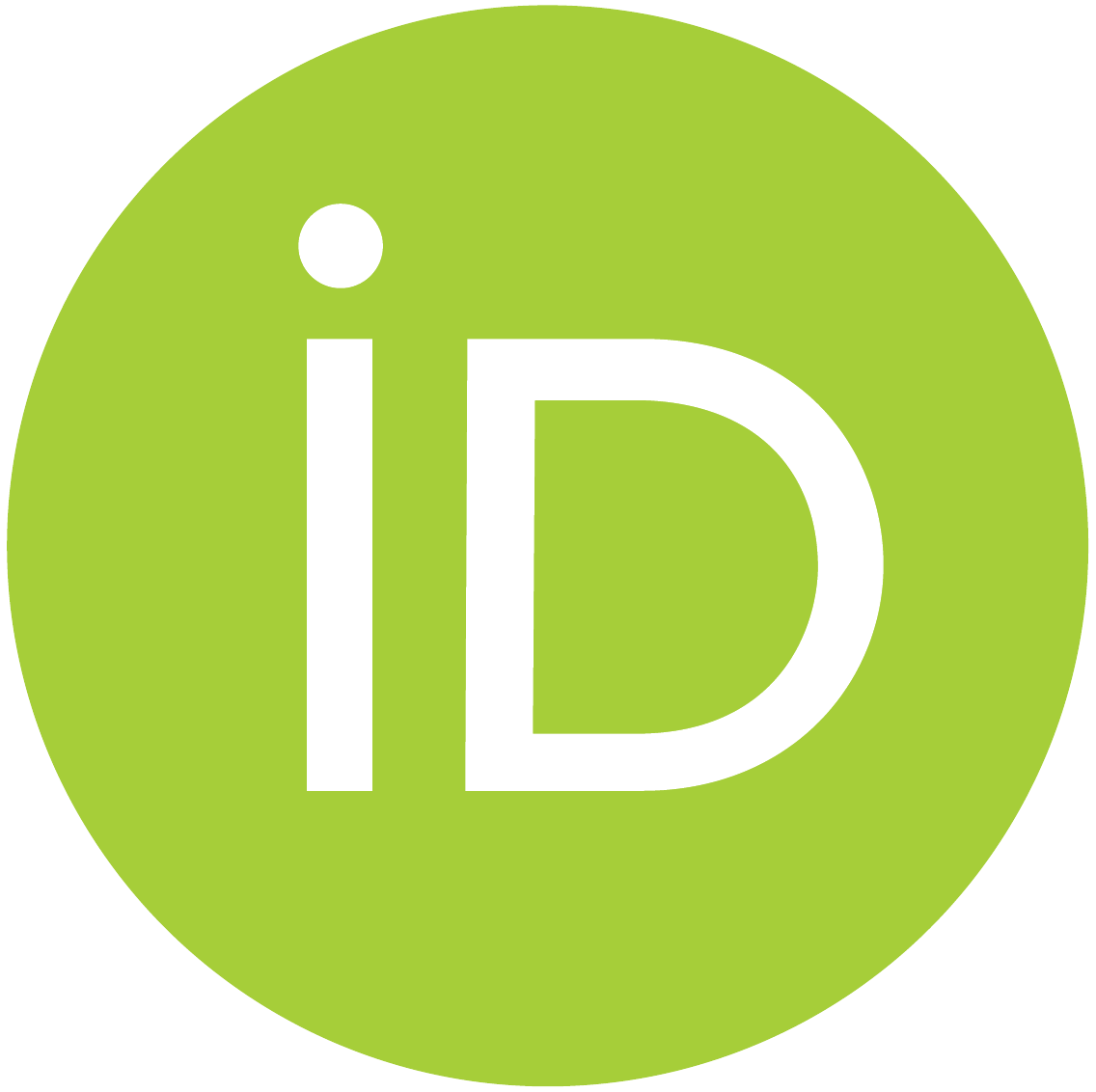}}

\newcommand{\orcid}[1]{\href{https://orcid.org/#1}{\orcidicon}}

\author{
Maik Ender\orcid{0000-0002-0685-2541}\inst{1} \and Felix Hahn\orcid{0009-0008-9260-4288}\inst{1}
\and Marc Fyrbiak
\orcid{0000-0002-4266-7108}\inst{1} \and\\ Amir Moradi\orcid{0000-0002-4032-7433}\inst{2} \and Christof Paar\orcid{0000-0001-8681-2277}\inst{1}
}

\institute{
Max Planck Institute for Security and Privacy, Bochum, Germany\\
\email[maik.ender@mpi-sp.org, felix.hahn@mpi-sp.org, christof.paar@mpi-sp.org]{firstname.lastname@mpi-sp.org}
\and
Technische Universität Darmstadt, Darmstadt, Germany\\ \email{amir.moradi@tu-darmstadt.de}
}

\authorrunning{Maik Ender, Felix Hahn, Marc Fyrbiak, Amir Moradi and Christof Paar}

  \title{JustSTART: How to Find an RSA Authentication Bypass on Xilinx UltraScale(+) with Fuzzing}

\titlerunning{JustSTART}

\begin{document}

\maketitle


\begin{abstract}
Fuzzing is a well-established technique in the software domain to uncover bugs and vulnerabilities.
Yet, applications of fuzzing for security vulnerabilities in hardware systems are scarce, as principal reasons are requirements for design information access, i.e., HDL source code. Moreover, observation of internal hardware state during runtime is typically an ineffective information source, as its documentation is often not publicly available.
In addition, such observation during runtime is also inefficient due to bandwidth-limited analysis interfaces, i.e., JTAG, and minimal introspection of hardware-internal modules.

In this work, we investigate fuzzing for Xilinx 7-Series and UltraScale(+) FPGA configuration engines, the control plane governing the (secure) bitstream configuration within the FPGA.
Our goal is to examine the effectiveness of fuzzing to analyze and document the opaque inner workings of FPGA configuration engines, with a primary emphasis on identifying security vulnerabilities.
Using only the publicly available hardware chip and dispersed documentation, we first design and implement \fwname, an advanced FPGA configuration engine fuzzing and rapid prototyping framework.
Based on our detailed understanding of the bitstream file format, we then systematically define 3 novel key fuzzing strategies for Xilinx FPGA configuration engines.
Moreover, our strategies are executed through mutational structure-aware fuzzers and incorporate various novel custom-tailored, FPGA-specific optimizations to reduce search space.
Our evaluation reveals previously undocumented behavior within the configuration engine, including critical findings such as system crashes leading to unresponsive states of the whole FPGA.
In addition, our investigations not only lead to the rediscovery of the recent starbleed attack but also uncover a novel unpatchable vulnerability, denoted as {JustSTART} (CVE-2023-20570), capable of circumventing RSA authentication for Xilinx UltraScale(+).
Note that we also discuss effective countermeasures by secure FPGA settings to prevent aforementioned attacks.
\end{abstract}

\section{Introduction}

\acp{FPGA} are a foundation in modern digital system landscape as their \emph{field-programmable} nature offers flexibility and adaptability.
To realize its flexibility, the \ac{FPGA} consists of 2 parts: i) the configuration engine that handles loading of a so-called \textit{bitstream} (representing a digital gate-level design) into ii) the \ac{FPGA} grid -- also called fabric -- that consists of millions of reconfigurable \acp{LUT} and routing to run the digital design, among other reconfigurable hardware primitives. 
Since \acp{FPGA} are commonly deployed in security-critical systems, including industrial control systems, cloud computing systems, and even military applications, manufacturers have implemented bitstream protection systems to ensure confidentiality, integrity, and authenticity.
While several works highlighted security vulnerabilities of the cryptographic implementation, i.e., via side-channel attacks~\cite{moradi2011vulnerability,swierczynski2014physical,moradi2016improved}, optical contactless probing~\cite{tajik2017power}, or implementation attacks~\cite{skorobogatov2012breakthrough, DBLP:conf/uss/Ender0P20, EnderLMP22, CVEXilinxUSHack}, a system-level methodical analysis of the configuration engine itself is -- to the best of our knowledge -- missing in the open literature.
However, since the configuration engine is complex (e.g., handling multiple security modes and various commands during the initialization process) and opaque (e.g., little to no information about its implementation details is publicly known, i.e., a gray-box setting), understanding its detailed architecture and security mechanisms plays an integral role for \ac{FPGA} security. Even more importantly, any vulnerability in the configuration engine is unlikely to be patched as it is  implemented in hardware and thus poses a massive threat.

In recent years, fuzzing in the software domain has seen widespread adoption in both academia and industry due to its effectiveness in uncovering software vulnerabilities. 
In particular, \textit{feedback-driven fuzzing}, i.e., generation of a random input mutation and observing system behavior that is then feed back into the input generation, has been shown to be effective. 
Even though various fuzzing methods have been proposed to analyze hardware systems~\cite{Trippel2021}, they leverage design information such as \ac{HDL} source code to perform its analysis and thus do not work for settings where only the manufactured chip is available with limited observable system information.

\paragraph{Goals and Contributions}
In this paper, we focus on fuzzing for Xilinx 7-Series and UltraScale(+) \ac{FPGA} configuration engines. Inspired by the aforementioned starbleed vulnerability in the configuration engine and capabilities of software fuzzing methods, our goal is to answer the following research question: 
\begin{quote}
    \textit{To what extent can we leverage systematic fuzzing techniques to derive (security) implementation information from the opaque configuration engine of Xilinx 7-Series and UltraScale(+) FPGAs?}
\end{quote}
In order to answer this research question, we first want to highlight that we -- as a security research community -- only have a superficial understanding of implementation details based on incomplete and dispersed publicly available limited documentation (gray-box setting). Note that this generally challenges effectiveness of fuzzing as interpretation of observed behavior is limited. In addition, effectiveness of fuzzing is another key challenge as we are generally limited by the device connection, e.g., via USB/\acs{JTAG}. From a high-level point of view, our goal is to increase the public knowledge about configuration engine implementation details with a focus on \emph{how} certain security functionalities are implemented.
Therefore, our approach and contribution are as follows 
\begin{itemize}
	\item{\bfseries \fwname FPGA Configuration Engine Fuzzing Framework (\autoref{js::section::framework})}
	We design and implement \fwname, a mutational bitstream fuzzing framework on the basis of a rapid prototyping approach to define bitstreams in a declarative manner.
	With context-specific \ac{FPGA} optimization, such as structure-aware bitstream mutations using a bitstream grammar and auto-generated encryption blocks, we improve the efficacy of our approach (that is generally limited by \acs{JTAG} speed).
    We publicly released \fwname under the MIT license on GitHub~\cite{RapidBitFuzzGithub}.

	\item{\bfseries Fuzzing Strategies (\autoref{js::chapter::strategies})}
    We develop three main strategies to systematically evaluate the 7-Series and UltraScale(+) series configuration engine: i) bitstream structure, ii) inter command, and iii) intra command. While the first strategy analyzes the general bitstream structure, the latter two concentrate on the commands.
    
	\item{\bfseries JustSTART (\autoref{js::chapter::case-studies::juststart})}
	Our evaluation uncovers a new unpatchable vulnerability named \textit{JustSTART} (CVE-2023-20570) that bypasses the \acs{RSA} authentication of Xilinx UltraScale(+) \acp{FPGA} and thus enables attackers to load trojanized or modified bitstreams. We want to note that this attack can be mitigated when both \ac{FPGA} bitstream encryption and authentication are enabled. 

	\item{\bfseries Further Security Vulnerabilities \& Understanding (\autoref{js::chapter::case-studies})}
	In further case studies on the 7-Series and UltraScale(+), we automatically uncover the recent starbleed attack and various undocumented behavior, for example, crashing the \ac{FPGA} into an unresponsive state or an undocumented \acs{RSA} test mode.
	We want to highlight that based on our findings, we also contribute to the understanding of the opaque configuration engine.
\end{itemize}

\section{Background \& Related Work}

To understand the mechanics of the Xilinx UltraScale(+) \acp{FPGA} configuration process, we now provide some fundamental \ac{FPGA} background aspects. Afterward, we detail related work focusing on hardware and embedded device fuzzing.

\subsection{Xilinx UltraScale(+) \texorpdfstring{\acp{FPGA}}{FPGAs}}\label{js::chapter::background::bitstream}
\acp{FPGA} are a class of \acp{IC} containing re-programmable logic to enable users to change a digital design even after manufacturing.
To this end, \acp{FPGA} consists of the so-called \textit{fabric}, i.e., grid of reconfigurable hardware elements such as \acp{LUT}, \acp{DSP}, \acp{BRAM}, I/O, and reconfigurable routing connecting all elements.
From a high-level perspective, the bitstream consists of a header to handle the configuration and the fabric configuration data containing the proprietary encoding of the gate-level digital design description. 
Since we are only interested in the configuration process, we refer the interested reader to diverse publications and open-source frameworks~\cite{prjxray, debit, ender2019insights, DingWZZ13, rannaud2008bitstream} that deal with the fabric configuration data.
We want to note that the header (starting with a sync word \texttt{0xAA995566}) consists of commands and data to read and write to the configuration engine registers.

\begin{figure*}[tbh]
    \centering
    \includegraphics[width=0.9\textwidth]{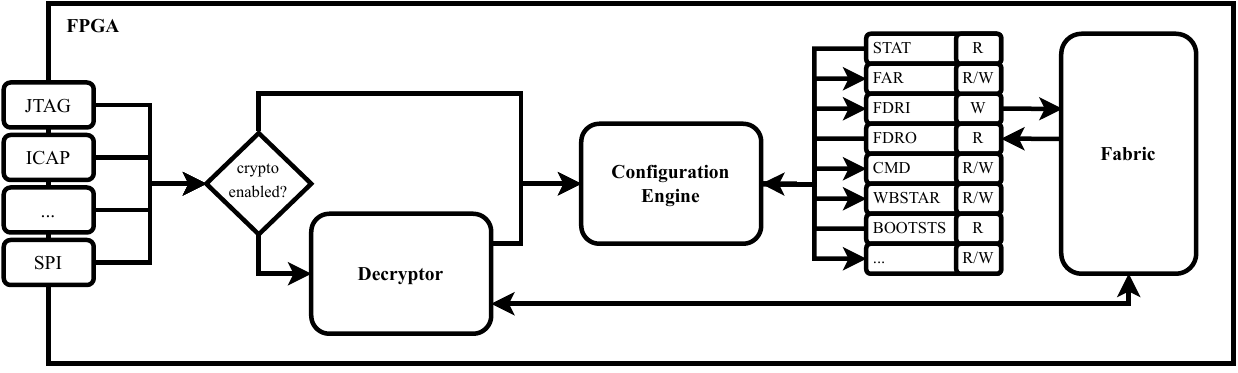}
    \caption{Our mental model of the configuration engine.}
    \label{js::fig::config_engine_model}
\end{figure*}

\paragraph{Configuration Process}\label{sec:configuration_process}
According to the user guide \emph{UG570}~\cite{UG570}, Xilinx UltraScale(+) \acp{FPGA} are configured by loading the application-specific configuration data into the internal memory of the fabric.
This configuration is done by loading the bitstream through the configuration engine via 1 of 5 possible configuration interfaces into the device, i.e., SelectMAP, \ac{ICAP}, \ac{SPI}, \ac{BPI}, or \ac{JTAG}.
Hence, this configuration engine manages the configuration process, which is crucial for \ac{FPGA} security.
As noted before, bitstreams can be encrypted, so they may require decryption first.
Even though the configuration engine details are not publicly known in detail, we have set up a mental model of our current understanding in~\autoref{js::fig::config_engine_model}.

\paragraph{Configuration Packets}\label{sec:configuration_packets}
Commands and data are organized as 32-bit words, where a packet header indicates a read, write, or NOP to a desired register and the number of words written to that register.
The data written to that register follows after the header.
\autoref{js::tab::type1_header} shows the type 1 packet header format. 
11 header bits are marked as \emph{reserved}, meaning they have no function and are reserved for future use.
Type 2 headers extend type 1 headers and lack the register address to maximize the bits used for the word count field.

\begin{table}[htbp]
\begin{center}
\resizebox{\columnwidth}{!}{
    \begin{tabular}{|c|c|c|c|c|}\hline
        \textbf{Header Type} & \textbf{Opcode} & \textbf{Register Address} & \textbf{Reserved} & \textbf{Word Count}\\\hline
        [31:29] & [28:27] & [26:13] & [12:11] & [10:0]\\\hline
        \texttt{001} & \texttt{xx} & \texttt{RRRRRRRRRxxxxx} & \texttt{RR} & \texttt{xxxxxxxxxxx}\\\hline
    \end{tabular}
}
\caption{
    Type 1 packet header: bits marked with \emph{R} have no functionality and are \emph{reserved} for future use (based on~\cite{UG570}).
}
\label{js::tab::type1_header}
\end{center}
\end{table}

\paragraph{Configuration Registers}\label{sec:config_registers}
The configuration engine uses the configuration registers to manage its internal state and configuration.
The only way to communicate with the configuration engine is to read and write to these registers.
For example, the \texttt{FDRI} register is used to write configuration data to the fabric of the \ac{FPGA}, and the \texttt{STAT} register contains information about the current status of the configuration engine.
5 bits address the registers as shown in \autoref{js::tab::type1_header}.
Hence, 32 registers are addressable at most.
According to the publicly available documentation, there are only 20 registers, leaving 12 undocumented registers.
In anticipation of our framework \fwname in \autoref{js::section::framework}, we analyze these undocumented registers since there are indications that a subset of registers have actual use.
Note that we refer to them as \emph{unknown registers} throughout the work at hand.
While most registers consume 32-bit data, they are of individual length, e.g., the \texttt{GCM-IV} register consumes 4 words (= 128-bit) of a \ac{GCM} \ac{IV}.
Many registers allow read access to aid debugging.
Moreover, readback of the fabric data is also possible if allowed by the security configuration.

\paragraph{\ac{FPGA} Bitstream Security}
As noted before, \ac{FPGA} bitstream protection schemes enable the security goals \textit{confidentiality} and \textit{authenticity}.
To ensure hardware design confidentiality, the bitstream fabric data can be encrypted and readback disallowed.
The bitstream authenticity ensures that the bitstream is not manipulated and no malicious design is executed on the \ac{FPGA}.
Note that this hinders the integration of hardware Trojans and other bitstream-level attacks.

For the UltraScale(+) \acp{FPGA}, Xilinx implemented an \ac{RSA} authentication and \ac{AES} encryption mechanisms, which can be used solely or combined to ensure the aforementioned security goals.
The \ac{AES} used in either the \ac{GCM} or \ac{CTR} mode ensures bitstream confidentiality.
If the \ac{AES} is used solely, it is used in the \ac{GCM} mode to implement bitstream authenticity.
Also, a proprietary Galois field-based checksum \textit{X-GHASH} is implemented to ensure the authenticity of blocks of 8 words within the bitstream~\cite{EnderLMP22}.
Besides, an \ac{RSA} authentication mechanism can be utilized to ensure authenticity.
It can be used with a plaintext bitstream or an \ac{AES}-encrypted bitstream.
For the latter, the \ac{AES} is operated in the \ac{CTR} mode without the \ac{GCM} and \textit{X-GHASH} authentication.
The \ac{AES} keys can be stored in a \ac{BBRAM} or burned to fuses.
Similarly, an \ac{SHA-3} hash of the \ac{RSA} public key is stored in fuses.
Two other fuses enforce that only encrypted and/or RSA-authenticated bitstreams can be loaded.
In summary, three different combinations of security measures can be used to ensure the authentication and integrity of the \ac{FPGA} bitstream: \ac{AES}-\ac{GCM}, \ac{AES}-\ac{RSA}, or plain-\ac{RSA}.

\subsection{Related Work}\label{sec:related_work}
We now provide a brief overview of existing literature encompassing \ac{FPGA} security.

\paragraph{\ac{FPGA} Attacks}
Since \acp{FPGA} are a foundation for many systems, they are a commonly targeted device.
Attacks against the configuration engine can be divided into 3 main areas:
i) side-channel attacks~\cite{Hettwer_Leger_Fennes_Gehrer_Gueneysu_2020, moradi2011vulnerability, swierczynski2014physical, moradi2016improved}  leverage information leaked through power consumption or electromagnetic radiation,
ii) probing attacks~\cite{tajik2017power, lohrke2018key} recover internal states or the key,
and lastly iii) implementation attacks~\cite{DBLP:conf/uss/Ender0P20, EnderLMP22, skorobogatov2012breakthrough, CVEXilinxUSHack} exploit vulnerabilities in the configuration engine implementation.

\paragraph{Hardware Fuzzing}
Hardware fuzzing is a subgroup of the fuzzing landscape.
Analogous to software fuzzing approaches, hardware fuzzing~\cite{Trippel2021, processorFuzz, le_detection_2019, Melotti2021, debnath_fuce_2021, DBLP:conf/nfm/PferscherA22, henderson_vdf_2017, xu_framework_2021, eisele_embedded_2021, qasem_automatic_2021, eceiza_fuzzing_2021, DBLP:conf/ndss/MuenchSKFB18} leverages the \ac{HDL} source code to observe the hardware behavior via simulation.
For example, in recent work, Canakci~\textit{et~al.}~\cite{processorFuzz} analyze an embedded processor implementation by monitoring the transitions in its control and status registers during simulation.
Similar to our work, they rely on registers of the target device to provide feedback and guide the fuzzing process.
However, in our approach, we do not have the \ac{HDL} source code of the configuration engine and cannot leverage any of the aforementioned hardware fuzzing approaches.
Methods for black-box or gray-box fuzzing on the hardware are scarce in the open literature. 
In 2010, Koscher~\textit{et~al.}~\cite{koscher_experimental_2010} analyzed embedded devices leveraging fuzzing, focusing the CAN bus in automotive vehicles. 
Since then, several works targeted embedded devices fuzzing with limited observable behavior~\cite{lee_fuzzing_2015, broek_security_2014, kamel_analysis_2013, alimi_analysis_2014, cheng_pdfuzzergen_2022}.
Additionally, fuzzers targeting the \ac{ISA} of processors have been proposed, for example~\cite{sandsifter} uncovers hidden instructions, glitches, and vulnerabilities in x86 \ac{ISA}, also aiming to understand the hardware better.

While such approaches have certain similar characteristics to our approach, e.g., no source code available and limited system behavior observation, key differences are the context in which fuzzing is carried out as the fuzzing methods are context-dependent and have to be adjusted for each scenario. 
Note that no prior work is dealing with fuzzing the \ac{FPGA} configuration engine.

In the recent work by Easdon et~al.~\cite{DBLP:conf/uss/Easdon0SG22}, a rapid prototyping framework is leveraged to develop attacks and identify vulnerabilities.
Similarly, we have adopted such rapid prototyping approach to extend our investigation and contribute to the general understanding of the configuration engine's inner workings and find vulnerabilities.

\section{\fwname Framework}\label{js::section::framework}

We now introduce the design and implementation of our \ac{FPGA} configuration engine fuzzing framework \fwname. First, we describe several key challenges for \ac{FPGA} configuration engine fuzzing, then we present the system architecture, workflow and fundamental components of \fwname~(\autoref{js::section::framework::architecture}), and finally provide implementation details~(\autoref{js::section::framework::implementation}).  

\paragraph{Key Challenges}
Since fuzzing of \ac{FPGA} configuration engines has the speed limitation of its underlying hardware, a key challenge is test cycle performance, i.e., bitstream generation and communication with the \ac{FPGA} for programming and subsequent information retrieval. A critical aspect of this challenge lies in generation of \textit{useful}, i.e., syntactically and semantically valid, test bitstreams to facilitate effective information retrieval within our gray-box setting. Note that without such structure-aware test generation, numerous test bitstreams may be generated that do not yield any useful (security) information as these are likely invalid (e.g., missing checksum values, invalid header commands, \dots).

\subsection{System Architecture}
\label{js::section::framework::architecture}
\begin{figure}[tb]
    \centering
    \includegraphics[width=\textwidth]{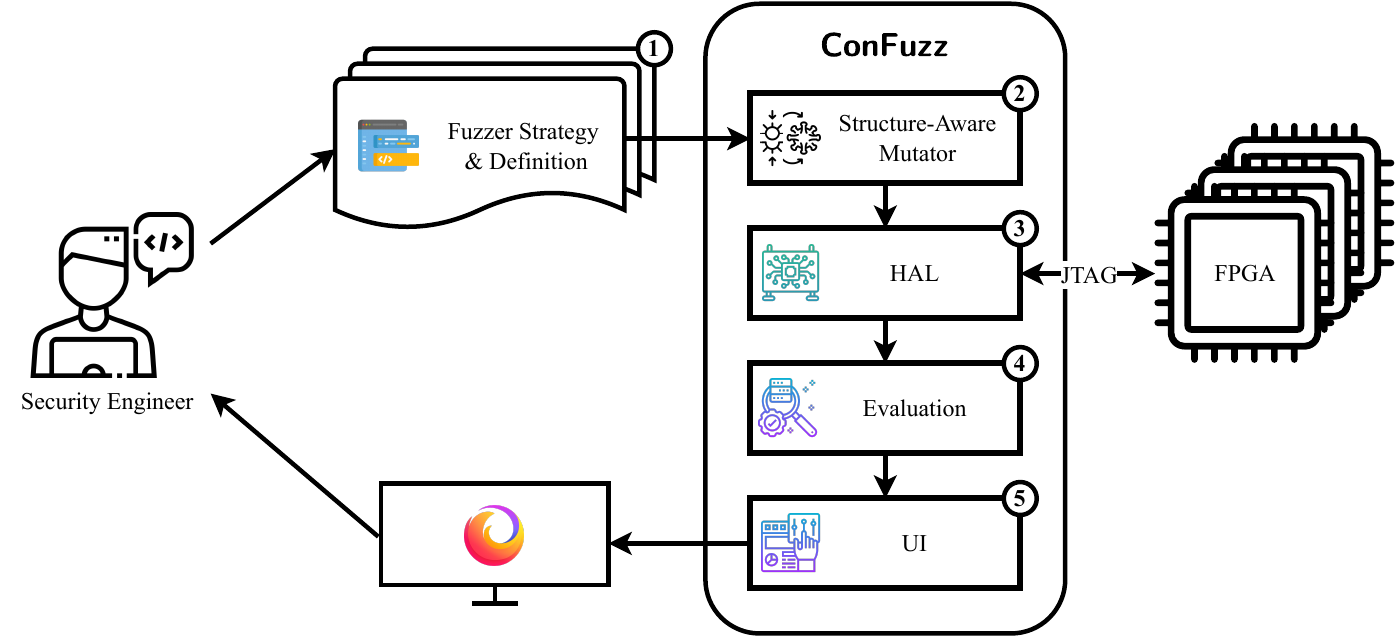}
    \caption{\fwname: Architecture and Workflow.}
    \label{js::figure::framework}
\end{figure}

\fwname is designed with high modularity and extensibility in mind. To this end, we structured \fwname upon several fundamental components, each characterized by a logically coherent feature set. In the following, we outline the workflow, providing more detail on the components, see \autoref{js::figure::framework}.

\paragraph{Workflow}
The user selects a fuzzing strategy and a target \ac{FPGA} to implement a concrete fuzzer~\circled{1}. The implemented fuzzer uses a structure-aware mutator~\circled{2} to generate modified test bitstreams automatically. Then, each modified bitstream is handled by the \acf{HAL}~\circled{3} in order to program the \ac{FPGA} and obtain information of the hardware state after configuration. Finally, the state is post-processed in an evaluator~\circled{4}, i.e., analyzed for crashes or deviated behavior, and optionally displayed in a user interface~\circled{5} for manual inspection.

To guide the reader, we now detail the components of \fwname in a bottom-up fashion, starting from the hardware to the user interface, as the higher-level components require context of the lower-level components. Moreover, note that we designed \fwname with the aforementioned challenges in mind.

\paragraph{Hardware Abstraction Layer (HAL) \circled{3}}
In order to instrument \ac{FPGA} device(s), we designed a hardware abstraction layer.
Key parts are communication using \ac{JTAG} or \ac{UART} and managing the \ac{FPGA} configuration engine state.
In particular, state management is vital to infer information for fuzzing feedback (e.g., hardware-internal register values, or whether the \ac{FPGA} responds at all, \dots). Note that we are in a gray-box setting and cannot inspect all hardware internals during configuration, so our goal is to infer as much direct and indirect information as possible: the Xilinx \ac{FPGA} configuration engine only communicates via its configuration registers, see \autoref{js::chapter::background::bitstream}, thus we opted to read-back of all configuration registers.

\paragraph{Structure-Aware Mutator \circled{2}}
As identified before, a key challenge is the generation of \textit{useful} bitstreams, i.e., the generation of syntactic and semantic valid bitstreams. 
We have implemented a grammar-based fuzzing approach within the structure-aware mutator, which relies on a custom-tailored \ac{FPGA} bitstream format grammar.
This grammar is based on the bitstream file format, incorporating all bitstream functions, ranging from simple headers and data value fields to automated generated checksums.
Key components are automated encryption and authentication primitives to ensure correct handling of a bitstream block to encrypt and/or authenticate them.
Within these bitstream primitives, \textit{fuzzing masks} can be defined to automatically generate variants of the defined bitstream.
For example, one can define a fuzzing mask to test 3 bits written to a register, then $2^3$ bitstream would be generated testing all possible bit combinations for that field, while checksums, encryption, etc. are handled automatically.

\paragraph{State Evaluation \circled{4}}
Fuzzing intends to find unexpected \textit{crashed} states. Therefore, upon the automated bitstream generation, transmission to the \ac{FPGA}, and subsequent retrieval of its state after programming, the state is evaluate  with careful analysis for unexpected states.
Inspired by software fuzzing practices, we integrated the concept of a \textit{crash} into our evaluation utilizing a range of crash settings.
These settings encompass various criteria such as presence or absence of expected values within designated registers, and deviations from default states.
For example, one approach is to probe the initial \textit{clean} state immediately following device restart, then after transmitting a bitstream, any deviations to the state are flagged as a crashes.
Note that the definition of a crash is user-defined, contingent upon the specific intentions during evaluation.

\paragraph{Fuzzer Strategy \& Definition \circled{1}}
Utilizing the previously mentioned bitstream grammar and crash settings, the security engineer implements a fuzzer strategy. Note that in our comprehensive case studies, we implemented 71 fuzzers, each adhering to the three principal strategies outlined in Chapter~\ref{js::chapter::strategies} and~\ref{js::chapter::case-studies}.

\paragraph{UI/Viewer for Inspection \circled{5}}
In the final phase, the security engineer inspects generated crashes to derive insights from the fuzzing process, aiming to pinpoint potential bugs and vulnerabilities while enhancing understanding of the configuration engine.
Results from the fuzzing process are accessible through a web interface, facilitating efficient manual analysis.
This interface displays the current fuzzing process and shows information about every logged test case.
It provides details such as the transmitted bitstream, current register values, and the meaning of individual register bits if publicly documented.
Additionally, the interface highlights the executed crash evaluation, i.e., it shows any test case and corresponding registers and their values involved in a crash.

\subsection{Implementation}
\label{js::section::framework::implementation}

We implemented key components such as the \ac{FPGA} bitstream grammar and fuzzing strategies of \fwname using Python, in particular, to enable rapid prototyping during experiments and manual inspection.
For \ac{FPGA} bitstream response state management, we used the boofuzz SQLite database.
As a fuzzing harness and viewer, we leveraged boofuzz~\cite{pereyda_boofuzz_2022} and for communication with the \ac{FPGA}, we used OpenOCD~\cite{OpenOCD}.

\paragraph{On Rapid Prototyping}\label{js::framework::RSA_TEST_MODE}
Natural language in the \ac{FPGA} user guides and documentation provided by the vendors 
has the implicit drawback of ambiguity and misinterpretation by the reader.
Thus, a key aspect during our design and implementation of \fwname was a horizontal rapid prototyping approach in order to rapidly verify or falsify certain assumptions about our mental model of the \ac{FPGA} configuration engine implementation details based on the documentation. 
In particular, such rapid prototyping capability is advantageous for sense-making of certain behaviors, i.e., device crashes.
Note that an observed state after configuration does not \textit{explain} its implementation details.
Typically, this is a manual sense-making task to link the documentation and observed outcome to a mental model of the presumed implementation, as, for example, \cite{xilinx_rsa_patent} states:

\begin{quote}
    \textit{The configuration data 314 is also encrypted if the DEC bit in the plaintext header is set. In an example implementation, if a} \texttt{TEST\_MODE} \textit{bit is asserted in the DLC, the configuration control circuit will expect 24 frames of configuration data. If the} \texttt{TEST\_MODE} \textit{bit is not asserted, then the configuration control circuit will write configuration data to the configuration memory until there is an indication that the end of the device has been reached.}~\cite{xilinx_rsa_patent}
\end{quote}
Based on this public Xilinx patent description, we then leveraged the rapid prototyping capability of \fwname to explore the \texttt{TEST\_MODE} bit. As the high-bits of the Decryption Length Counter (DLC) are typically 0 for standard lengths, we assume that the higher bits are used for the \texttt{TEST\_MODE}. We then create a bitstream with the highest bit set to '1' and send 24 frames of configuration data. The bitstream was successfully programmed and thus revealed the presence of the \texttt{TEST\_MODE} bit in commercial-of-the-shelf Xilinx UltraScale(+) \acp{FPGA}. Note that we leverage the \texttt{TEST\_MODE} for optimization in our second case study (\autoref{js::chapter::case-studies::juststart}) as programming of whole fabric configuration data requires typically $\sim20$s for UltraScale(+) \acp{FPGA} while bitstreams with just 24 frames of configuration data are programmed in $\sim2$s, thus achieving a performance speed-up of $\sim10\times$.

\section{On Bitstream Fuzzing Strategies}\label{js::chapter::strategies}

It is crucial to acknowledge that our endeavor has its limitations.
The speed constraints imposed by the use of hardware, i.e., the \ac{JTAG} interface, prevent us from exhaustively fuzzing every possible configuration parameter.
So we have to take a strategic \emph{guided} approach to reduce the fuzzing test cases and test only \emph{promising} ones.
The available documentation at hand gives us a superficial understanding of the configuration engine such that we can build a first mental model of the configuration engine and bitstream grammar (cf. Section~\ref{js::chapter::background::bitstream}).
Based on this information, we define three main strategies: i)~bitstream structure, ii)~inter command, and iii)~intra command.
While the first strategy challenges the general bitstream structure, the latter two concentrate on the commands themselves. Here, the inter-command strategy tests explicitly the behavior of each single command, while the intra-command strategy examines the interaction between them.

Note that the purpose of this work is to better understand the undocumented part of the configuration engine and thus find potential vulnerabilities. The first two strategies are primarily designed to uncover undocumented parts within the bitstream and enhance our mental model of the configuration engine, while the third strategy is also intended for detecting security vulnerabilities by especially targeting the interaction with the bitstream protection schemes. In the following, we discuss these strategies in detail, followed by the general implementation procedure of a fuzzer.

\paragraph{Fuzzer Strategy 1: Bitstream Structure}

Xilinx documents the bitstream structure, as discussed in the background in Section~\ref{js::chapter::background::bitstream}. The configuration engine is generally organized in registers while the bitstream reads and writes to them. The bitstream is structured in header commands followed by the data and commands to be written in the specified registers (cf. Table~\ref{js::tab::type1_header}).

Within this first fuzzer strategy, we target this bitstream structure in order to uncover undocumented bits and unused bit combinations.
For example, the first three header bits determine the header type, e.g., 8 possible types could exist, while only two types are documented. Similarly, one unused opcode (11) exists, and eleven reserved bits.

\paragraph{Fuzzer Strategy 2: Intra Command}
The next layer of abstraction is configuration engine registers. The \ac{FPGA} configuration engine has (most likely~\footnote{
Table~\ref{js::tab::type1_header} shows that 9 of 14 address bits are marked as reserved. With the help of a strategy 1 fuzzer, we could not find any register influencing the \ac{FPGA} state, which is addressed via these 9 reserved address register bits. This concludes that only the 5 lower bits are used to address registers, hence a maximum of 32 addressable registers.}) 
a total of 32 registers; however, only 20 of these registers have been documented in public resources, while several bits are still marked as reserved even when bitstreams generated by Vivado contain them.

Within the second fuzzer strategy (intra-command), we specifically target one of each register at a time to uncover undocumented registers, document the bit usage, and test their influence on the \ac{FPGA} state.
For example, we created a default intra-register fuzzer for the several completely undocumented registers.
This fuzzer writes a single 32-bit data word to the fuzzed register, as the typical registers are of 32-bit size.
We provide a detailed discussion of a fuzzer that implements the intra-command strategy in our latest case study presented in Section~\ref{js::chapter::case-studies::unknwon_23}.

\paragraph{Fuzzer Strategy 3: Inter Command}
Moving beyond the analysis of the bitstream structure and individual commands, we explore the interactions between multiple commands and their potential ramifications within our last fuzzing strategy.
One particular aspect of our inter-command fuzzing strategy involves investigating the bitstream protection features.
For example, commands are placed around the new RSA authentication mode of UltraScale(+) devices.
We provide a detailed discussion of a fuzzer that implements the inter-command strategy in our second case study presented in Section~\ref{js::chapter::case-studies::juststart}.

\paragraph{General Fuzzer Implementation Procedure}
A hallmark of our approach is the systematic procedure we can follow for the implementation of each fuzzer, regardless of the chosen strategy. The general workflow comprises the following steps:

\begin{enumerate}
    \item \textbf{Selection of Target/Strategy:} First, target \ac{FPGA} and board are selected, as well as one of the main fuzzing strategies. Based on public documentation, we can make assumptions regarding the configuration engine to pick a specific target within the configuration engine aligned with the chosen strategy.
    This enables systematically covering all areas of the configuration engine within a specific strategy.
    
    \item \textbf{Fuzzer Construction:} With the selected fuzzing strategy and target in mind, a specific fuzzer is implemented. 

    \item \textbf{Fuzzer Execution \& Crash Logging:} The fuzzer is then executed, automatically generating bitstreams, sending them to an FPGA, and probing the \ac{FPGA} state. Only \textit{abnormal} states are logged as a crash depending on user-defined crash settings.
    
    \item \textbf{Interpretation and Refinement:}
    Identified crashes are then analyzed to discern their underlying causes and extend our mental model of the bitstream configuration engine. This process of interpretation informs the iterative refinement of the fuzzer and its parameters. In certain scenarios, we even employ secondary fuzzers designed to explore specific bit positions, configuration parameters, and crash settings or utilize the rapid prototyping feature of \fwname.

\end{enumerate}

\section{Case Studies}\label{js::chapter::case-studies}
With the fuzzing strategies from \autoref{js::chapter::strategies} in mind, we implemented 71 fuzzers on Xilinx 7-Series and UltraScale(+) FPGAs.
As we cannot discuss every finding in detail in the scope of this work, we provide an excerpt of the most interesting and impactful findings in \autoref{js::table::fuzzer_results}; a complete list of all findings is available at~\cite{RapidBitFuzzGithub}. Three findings are discussed in this chapter, showcasing the implementation and investigation procedure.

With the $71$ implemented fuzzers, we executed about $83$ million test cases, which took approximately 2 weeks.
For UltraScale(+) fuzzers, we used a single board and a small cluster of 15 \acp{FPGA} for the Xilinx 7-Series.
In our experiments, we tested up to $2^{21}$ on UltraScale(+) and on our 7-Series cluster $2^{25}$  bitstreams per day, depending on the fuzzer configuration.
We found $1677$ crashes during evaluation.
Overall, our approach facilitated an improved understanding of the bitstream file format and implementation of the configuration engine, including errors and crashes, see \autoref{js::table::fuzzer_results}.

\begin{table}[ht]
    \centering
    \begin{tabular}{lp{1.5cm}lp{7cm}}
        \toprule
        \textbf{Name} & \textbf{FPGA Family} & \textbf{Type} & \textbf{Findings} \\
        \midrule
        header types & US+ & structure & Header type 010 always and header type 001 with opcode 11 does lead to a crash of the device (all registers are zero) \\\hline
        register 23 & 7S, US+ & intra & Bit 23 (\& 25) crashes the device (power cycle needed) \\\hline
        register 29 & 7S, US+ & intra & Value matches FUSE\_CNTL register (documented JTAG only) \\\hline
        starbleed & 7S, US+ & inter & Re-discovers Starbleed attack~\cite{EnderLMP22} \\\hline
        JustSTART & US+ & inter & Discovers novel JustSTART attack \\
        \bottomrule
    \end{tabular}
    \caption{Excerpt of our fuzzers and findings (7S: 7-Series, US+: UltraScale(+)).}
    \label{js::table::fuzzer_results}
\end{table}

\paragraph{Development Boards}
For our experiments, we examined the following Xilinx \acp{FPGA} development boards:
Basys 3 (Artix-7 XC7A35T), KCU116 (UltraScale+ Kintex XCKU5P), and OpalKelly XEM8320 (UltraScale+ Artix XCAU25P), while we also support the KCU105 (UltraScale Kintex XCKU040).
All boards offer \ac{JTAG}-over-USB for \ac{FPGA} programming.
Further, the jumpers are set to \emph{\ac{JTAG}} boot mode to prevent the \acp{FPGA} from being configured from any other bitstream source until a \ac{JTAG} programming occurs.

\subsection{Fuzzing Encrypted Bitstreams: How to Find Starbleed}\label{js::chapter::case-studies::starbleed}
The recent starbleed attack~\cite{DBLP:conf/uss/Ender0P20, EnderLMP22} is a time-of-check-to-time-of-use vulnerability where bitstream commands can be executed before verifying their authenticity.
Ender~\textit{et~al.} discovered the attack ``by a detailed study of the Xilinx official documents together with experiments'' which is a manual laborious task.
We employed an automated fuzzer, as outlined in this case study, to automate the process, leading to the rediscovery of the attack. Moreover, we found 3 additional registers vulnerable to the attack.
Note that this case study focuses on UltraScale(+) devices.
The fuzzer for the 7-Series works analogously, and results are available in our GitHub repository~\cite{RapidBitFuzzGithub}.
We further assume that no \ac{RSA} authentication is enforced by the security fuses, as it would prevent the attack.

\paragraph{The Starbleed Attack}
The root cause of starbleed is that manipulations of the ciphertext are detected after decryption and execution of commands.
A manipulation can be inserted by flipping bits in an encrypted ciphertext undermining the used \ac{GCM} mode and the underlying \ac{CTR} mode malleability.
An attacker can manipulate the bitstream to divert secret fabric content to the \texttt{WBSTAR} register instead. 
Then, when the \ac{FPGA} finally detects the manipulation, the \ac{FPGA} resets but does not clear said register.
So it can be read out afterward.
This behavior can be exploited to read out all confidential data within the bitstream via the \texttt{WBSTAR} register.
For example, \autoref{js::fig::starbleed::bitstream_structure} shows parts of an encrypted bitstream.
$\beta_8$ is the first encrypted bitstream command.
Manipulating this word would divert the following data words.

\paragraph{The Starbleed Fuzzers Idea}
In the following, we assume a scenario where the analyst has no prior knowledge about the starbleed attack but has a basic understanding of the bitstream structure from the available documentation.
To detect undefined behavior, the analyst starts fuzzing an encrypted bitstream like an unencrypted one by starting with a strategy 1 fuzzer.
This strategy tests the bitstream structure, e.g., the header words of the bitstream are fuzzed.
We designed the fuzzer to cover all 21 active bits in the header (see~\autoref{js::tab::type1_header}), i.e., we omitted the reserved ones.
We exclude the reserved bits in this particular fuzzer as the mutation space is too large, and we have already tested these bits with an unencrypted structural fuzzer, which revealed that they likely have no functionality.

In order to mimic an attacker capability who can only flip encrypted bits instead of generating correctly encrypted bitstreams, we directly mutated an encrypted bitstream rather than mutate and then encrypt the bitstream.
Note that this manipulates only one word at a time, e.g., $\beta_8$ in \autoref{js::fig::starbleed::bitstream_structure}; no other words are changed, especially checksums and authentication tags.
We created a short encrypted bitstream writing 3 frames to the fabric to reduce the fuzzed bitstream size and improve the fuzzer performance.
The implementation of the starbleed fuzzer is displayed in \autoref{js::lst::starbleedrequest}.

\begin{figure*}[tb]
\centering
\begin{bytefield}[bitwidth=0.205\textwidth,bitheight=3em]{4}
	\bitheader{0-3} \\
	
 
	\begin{rightwordgroup}{\rotatebox{90}{\parbox{1cm}{unencrypted\\configuration\\header}}}
		\bitbox{1}{\footnotesize 0xAA995566\\ SYNC Word $\beta_{0}$} &
        \bitbox{1}{\footnotesize 0x20000000\\ NOP $\beta_{1}$} &
        \bitbox{1}{\dots} &
        \bitbox{1}{\footnotesize 0x30016004\\ write GCM IV $\beta_{3}$} \\
	
		\bitbox{3}{\footnotesize GCM IV 96 bit\\ $\beta_{4}$ --  $\beta_{6}$} &
        \bitbox{1}{\footnotesize enc length\\ $\beta_{7}$} \\
        
		\bitbox{4}{60 * NOPs (0x20000000)}
	\end{rightwordgroup} \\
	
	\begin{rightwordgroup}{\rotatebox{90}{\parbox{3cm}{encrypted part\\ shown in plaintext}}}
			\bitbox{1}{\footnotesize 0x244A2E44\\ checksum $\gamma_0$} &
            \bitbox{1}{\footnotesize \color{red} 0x30018001\\ write IDCODE $\beta_{8}$} &
            \bitbox{1}{\footnotesize 0x03822093\\ IDCODE value $\beta_{9}$} &
            \bitbox{1}{\footnotesize 0x30002001\\ write FAR $\beta_{10}$} \\
            
			\bitbox{1}{\footnotesize 0x00000000\\ frame address $\beta_{11}$} &
            \bitbox{1}{\footnotesize 0x30008001\\ write CMD $\beta_{12}$} &
            \bitbox{1}{\footnotesize 0x00000001\\ WCFG command $\beta_{13}$} &
            \bitbox{1}{\footnotesize 0x30004000\\ write FDRI (type 1) $\beta_{14}$} \\
            
			\bitbox{1}{\footnotesize 0x0BB5E100\\ checksum $\gamma_1$} &
            \bitbox{1}{\footnotesize 0x50000171\\ write FDRI (type 2) $\beta_{15}$} &
            \bitbox{1}{\footnotesize 0xF00DF00D\\ fabric data $\beta_{16}$} &
            \bitbox{1}{\footnotesize 0xF00DF00D\\ fabric data $\beta_{17}$} \\
            
		
			\bitbox{1}{\footnotesize 0xF6012B13\\ checksum $\gamma_2$} &
            \bitbox{1}{fabric data} &
            \bitbox{1}{fabric data} &
            \bitbox{1}{\dots} \\
            
			\bitbox{4}{\dots}

  
	
	\end{rightwordgroup} \\
 
\end{bytefield}
\caption{The encrypted bitstream utilized to be fuzzed by the starbleed fuzzer.}
\label{js::fig::starbleed::bitstream_structure}
\end{figure*}

\paragraph{Fuzzer Execution \& Analysis}
On a single OpalKelly XEM8320 Ultrascale+ board, the starbleed fuzzer stops execution after recording 128 crashes by default, having executed $20,119,674$ test cases in roughly 8 days.
All crashes occur while mutating $\beta_15$ and directly reveal the starbleed vulnerability, as one can see decrypted fabric data in configuration registers.
While we tried to keep the fuzzer configuration as simple as possible, resulting in an extended runtime of the fuzzer, it is plausible that a knowledgeable attacker could refine the fuzzer strategy to find the vulnerability more quickly, e.g., by modifying the starting position of the first fuzzed word or relaxing the crash settings to match other decrypted data.

On the 7-series cluster, using 15 boards in parallel, some boards fuzzed a test case range where no crashes occurred, while other boards exited early after recording 128 crashes. 
This results in a runtime of about 46 hours with $43,189,976$ test cases and 1156 crashes.
Most crashes are similar to each other, as, for example, the same manipulations are carried out at different words in the bitstream.
Again, most of them directly revealed the starbleed vulnerability, while some are not linked to starbleed, as data words are manipulated and not header words.
Note that the results of the UltraScale+ starbleed fuzzer and the starbleed fuzzer on the 7-series cluster are not directly comparable due to subtle implementation differences.

With these results, the analyst would have discovered the starbleed vulnerability, i.e., manipulating an encrypted (non-RSA-authenticated) bitstream is possible as the bitstream commands are executed before the next cryptographic checksum is validated.
However, the X-GHASH checksum is still an issue within the practicability of that attack because this checksum ($\gamma_i$) hinders leakage of bitstream content as it is evaluated every 8\textsuperscript{th} word.
Nevertheless, the underlying attack foundation is found at this stage.

\paragraph{New Discoveries}
Through our fuzzing attempts, we have discovered that in addition to the \texttt{WBSTAR} register, the \texttt{TIMER}, \texttt{UNKNOWN\_20}, and \texttt{BSPI} registers are also susceptible to the starbleed attack.
To the best of our knowledge, this was not publicly known yet.

Accordingly, on the Virtex-6, the starbleed attack might become usable again as the \texttt{WBSTAR} register lacks the two most significant bits, i.e., the \texttt{WBSTAR} register is only 30 bits long.
Unfortunately, we could not test this yet due to the lack of a legacy Virtex-6 board.

\subsection{JustSTART: \texorpdfstring{\ac{RSA}}{RSA} Authentication Bypass}\label{js::chapter::case-studies::juststart}
As of now, \emph{no} vulnerability to the \ac{RSA} bitstream authentication of Xilinx UltraScale(+) \acp{FPGA} was publicly known.
Unfortunately, we discovered a flaw in said \ac{RSA} authentication mechanism of Xilinx UltraScale(+) \acp{FPGA}.
With the JustSTART vulnerability, the \ac{RSA} bitstream is loaded to the \ac{FPGA} as usual, but instead of running the \ac{RSA} authentication mechanism, commands to \emph{just start} the \ac{FPGA} are inserted into the bitstream, and the device boots successfully.
We discovered this flaw using our fuzzing framework \fwname.
With this vulnerability, the \ac{RSA} authentication of all Xilinx UltraScale(+) \acp{FPGA} can be bypassed, even if \ac{RSA} is enforced by the security fuses.
Currently, the only countermeasure to this attack is to enforce the bitstream encryption by the \emph{\ac{AES} only} fuse since the device does not decrypt the attacking bitstream.
In the following, we describe the attack in detail and its consequences, followed by a description of finding it with \fwname.

\subsubsection{\texorpdfstring{\ac{RSA}}{RSA} Authentication for UltraScale(+)}
\begin{figure}[tbh]
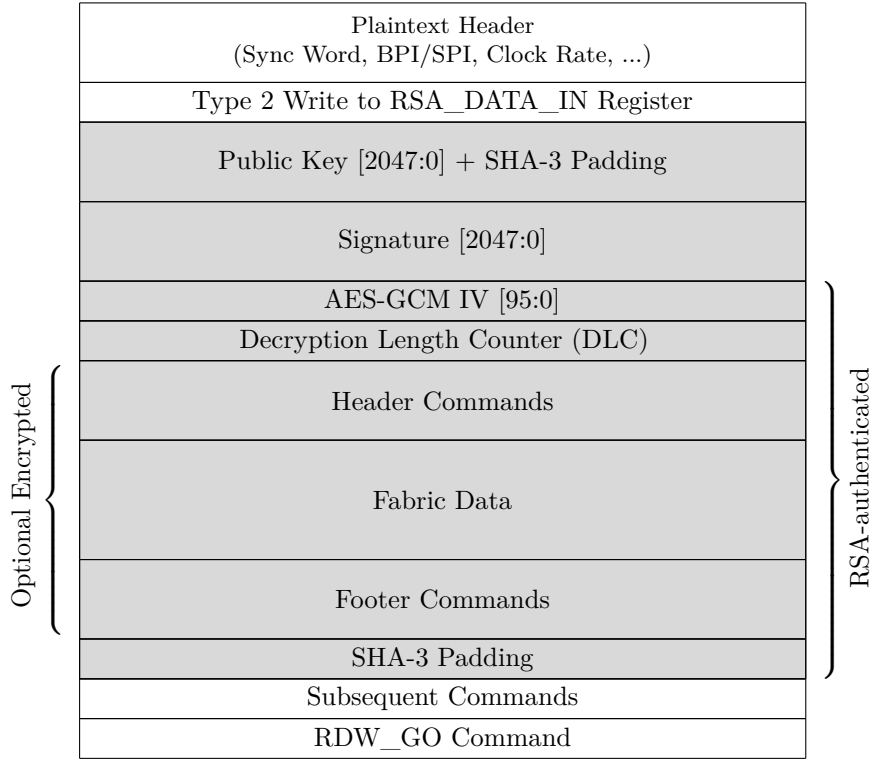

\centering
\begin{bytefield}[bitwidth=0.7\textwidth,bitheight=1.5em]{1}
	\small
	\wordbox{2}{Plaintext Header\\(Sync Word, BPI/SPI, Clock Rate, ...)} \\
	\wordbox{1}{Type 2 Write to RSA\_DATA\_IN Register} \\
    \wordbox{2}[bgcolor=gray_mild]{Public Key [2047:0] + SHA-3 Padding}\\
    \wordbox{2}[bgcolor=gray_mild]{Signature [2047:0]}\\
    \begin{rightwordgroup}{\rotatebox{90}{\parbox{3cm}{RSA-authenticated}}}
        \wordbox{1}[bgcolor=gray_mild]{AES-GCM IV [95:0]}\\
        \wordbox{1}[bgcolor=gray_mild]{Decryption Length Counter (DLC)}\\
        \begin{leftwordgroup}{\rotatebox{90}{\parbox{3cm}{Optional Encrypted}}}
            \wordbox{2}[bgcolor=gray_mild]{Header Commands}\\
            \wordbox{3}[bgcolor=gray_mild]{Fabric Data}\\
            \wordbox{2}[bgcolor=gray_mild]{Footer Commands}
        \end{leftwordgroup}\\
    \wordbox{1}[bgcolor=gray_mild]{SHA-3 Padding}
    \end{rightwordgroup}\\
    \wordbox{1}{Subsequent Commands}\\
    \wordbox{1}{RDW\_GO Command}
\end{bytefield}
	\caption{RSA Bitstream Structure based on~\cite{XAPP1098}. Gray-shaded words are written to the \texttt{RSA\_DATA\_IN} Register.}
	\label{js::fig::rsa_bitstream_structure}
\end{figure}

\autoref{js::fig::rsa_bitstream_structure} describes the structure of an \ac{RSA}-authenticated bitstream~\cite{UG570, XAPP1098, xilinx_rsa_patent}.
The bitstream starts with an unauthenticated plaintext header followed by a write to the \texttt{RSA\_DATA\_IN} register.
The data written to this register includes the \ac{RSA} public key, \ac{SHA-3} padding, and \ac{RSA} signature, calculated over the following written data, the encryption IV, decryption length counter, header commands, fabric data, and footer commands.
Note here that the \ac{RSA} bitstream structure is very rigid, as the length of each written block is fixed.
Even the word count for the write to the \texttt{RSA\_DATA\_IN} is fixed depending on the \ac{FPGA} fabric size.
The following bitstream data, which is not authenticated and not written to the \texttt{RSA\_DATA\_IN} register, is the \texttt{RDW\_GO} and other commands at the bitstream footer.
Note that we discovered a test mode to break out of this rigid structure (see~\autoref{js::framework::RSA_TEST_MODE}).

When using an RSA-authenticated bitstream, its authenticity is verified before activating the decryption core and decrypting it.
This prevents time-of-check-to-time-of-use attacks and side-channel attacks to recover the decryption key.
For this reason, the authenticated bitstream data must be buffered on the device.
While the header and footer commands have a designated buffer, the fabric data is buffered in the \ac{FPGA} fabric, as it would consume too much space otherwise.
Hence, the write to the \texttt{RSA\_DATA\_IN} are writes to the header/footer buffers, as well as, the \ac{FPGA} fabric.
No commands are executed while writing to these buffers.
The first executed bitstream commands after buffering are the subsequent unauthenticated footer commands.
Then, the \texttt{RDW\_GO} (Read-decrypt-write\_GO) command initiates the authentication and decryption of the previously buffered data.
At first, the authenticity of the bitstream is checked.
If the authenticity is confirmed and the bitstream is not encrypted, the \ac{RSA} header and footer are executed, starting the \ac{FPGA} design.
If the bitstream is encrypted, the \ac{RSA} header is decrypted and executed.
Afterward, the fabric is decrypted frame by frame, i.e., read from the fabric, decrypted, and written back to its place in the fabric.
Finally, the \ac{RSA} footer is decrypted and executed.
This footer contains the commands to boot the device and run the design previously written to the fabric.
In order to complete these complex tasks of authenticating and decrypting the buffered bitstream, the \texttt{RDW\_GO} command needs many ticks on the configuration clock~\cite{UG570}.

\subsubsection{Attack Idea}
The idea of JustSTART is simple: replace the \texttt{RDW\_GO} command at the end of the bitstream with the usual start-up command sequence.
As a result, the authentication process is not started, and the design already present in the fabric is executed.

The attack works because the fabric is used as a buffer for the design, and the end of the bitstream, including \texttt{RDW\_GO} command, is not authenticated.
In the case of plaintext \ac{RSA} bitstreams, the design is already present in the fabric, waiting to be authenticated by the \texttt{RDW\_GO} command.
If an encrypted bitstream was loaded, replacing the \texttt{RDW\_GO} command prevents the activation of the decryption engine, leaving fabric data encrypted.
Therefore, the attack only works if the attacker bitstream is unencrypted.

\subsubsection{Attacker Model}
The presented attack allows an attacker to circumvent the \ac{RSA} authentication of Xilinx UltraScale(+) \acp{FPGA}.
By bypassing the authentication step and \emph{just starting} the device after the bitstream has been written to the \ac{FPGA} fabric, an attacker can configure the device with arbitrary plaintext \ac{RSA} bitstreams.
Hence, our attacker model is the following:
\begin{itemize}
    \item{\bfseries Bitstream Manipulation} The attacker needs to manipulate and load the bitstream to the attacked device by manipulating the non-volatile memory, storing the bitstream, or loading it via some configuration port.
    \item {\bfseries Non-Encrypted Attacker Bitstream} The bitstream loaded to the \ac{FPGA} by the attacker must not be encrypted as the \texttt{RDW\_GO} command, which triggers the decryption, is replaced.
    \item {\bfseries Fuses} Setting the \emph{\ac{AES} only} fuse, allowing only encrypted bitstreams, prevents the attack.
    However, enabling the \emph{\ac{RSA} only} fuse, which solely permits \ac{RSA}-authenticated bitstreams, does not prevent the attack.
\end{itemize}

\subsubsection{Discovering the Attack by Fuzzing}
We found this attack 3 times during our fuzzing experiments by applying different fuzzers.
We describe the most straightforward one in the following, shown in \autoref{js::lst::juststart_fuzzer}.
With this fuzzer, we inserted 3 writes to the command register at the unauthenticated end of a plaintext \ac{RSA} bitstream.
We chose said register as these commands directly influence the \ac{FPGA} state and play an immersive part in the startup sequence.
The fuzzer is configured to use a different \ac{RSA} public key than written to the fuses, resulting in an invalid \ac{RSA} signature in the bitstream.
Hence, if the \emph{\ac{RSA} only} fuse is set, the device should not be able to start.
Due to the impact of the command register, often influencing the device state, we narrowed down the crash settings.
To this end, we focused on identifying instances where the configuration engine does not report any errors or indicates a successful device startup in the \texttt{STAT} or \texttt{BOOTSTS} register.

\paragraph{Fuzzer Results}
The fuzzer processed $32,768$ test cases in 17 hours.
As stated previously, for valid \ac{RSA} bitstreams to start, \ac{JTAG} needs to execute a large number of cycles to process the \texttt{RDW\_GO} command, increasing the runtime of this fuzzer.
Note that we ran the tests on a KCU116 board with the \emph{\ac{RSA} only} fuse enabled.

Only 2 test cases are logged as crashed, where the device does not show any error, and the done pin is high, indicating a successful start.
In both test cases, the sent bitstreams are mostly the same, except for 2 commands that have been rearranged.
Investigating these commands reveals that they are part of a regular startup sequence to boot up the device.
Usually, such a startup sequence is present in the bitstream footer.
In order to verify the attack, we modified a valid design by invalidating the \ac{RSA} signature and replacing the \texttt{RDW\_GO} command with a valid startup sequence.

\subsubsection{Extending the Attack}\label{js::chapter::case-studies::juststart::extending_attack}
We also tried to extend this attack to break confidentiality, i.e., the encryption.
While we tried several fuzzing and rapid prototyping ideas to enable the decryptor, extending the attack remains unsuccessful.
Our main idea to extend the attack is to load an \ac{ICAP} controller with the JustSTART attack, then decrypt a short bitstream and read the decrypted content back via \ac{ICAP}.
The \ac{ICAP} interface is an internal configuration port considered trusted and thus lacks most security features.
Reading back any fabric content via \ac{ICAP} is allowed even when encrypted bitstreams are loaded.

While it is possible to load such an \ac{ICAP} controller with our attack, we remain unsuccessful in decrypting a previously encrypted bitstream.
The issue is two-fold.
First, with fuzzing, we tried to activate the decryption bit in the \texttt{CTL0} register, i.e., enable the decryptor after a plaintext \ac{RSA} bitstream is loaded, which seems impossible.
Second, if the decryption bit is enabled, running a decryption using a non-\ac{RSA}-authenticated encrypted bitstream is impossible, as we confirmed by rapid prototyping.
Already the documentation states~\cite{XAPP1098} that if \ac{RSA}-authenticated bitstreams are used, partial reconfiguration is only allowed via \ac{ICAP} with unencrypted and unauthenticated bitstreams.

\subsection{Investigating Unknown Register 23}\label{js::chapter::case-studies::unknwon_23}
In this case study, we investigate the unknown register 23.
For that, we used our default intra-register fuzzer and further investigated the crashes by means of rapid prototyping.
We uncovered some bits leading to an unresponsive state of the \ac{FPGA}, which can only be resolved by a power cycle.
These experiments were carried out on the UltraScale(+) with an \ac{AES} key programmed to the \ac{BBRAM}.
7-Series \acp{FPGA} are also sensitive to different bit combinations' nuances.

\paragraph{Intra Register Fuzzer}
To investigate an unknown register, we choose the intra-register strategy (cf. \autoref{js::chapter::strategies}) with a broad fuzzing mask by default.
This ensures that we get as much information as possible about the unknown register.
Since register 23 is unknown, e.g., it is not documented, and there are no bitstreams we could generate containing a write to the register, no individual bits or bit pattern are known.
This said we wanted to test most of all possible values the register could hold while respecting the speed limitation of fuzzing hardware.
Therefore, we split the 32-bit register into two 18-bit fuzzing masks to test.
The two masks overlap in 2 bits, e.g., bit patterns in that region are likely to be caught while reducing the fuzzing time.
Hence, this default intra-register fuzzer generates and tests $2^{19}$ bitstreams.

\paragraph{Investigating the Crashes}

When executing this default intra-register fuzzer for the unknown register 23 without defining register-specific crash settings, every test case is logged as crashed, and the fuzzer exits after 128 crashed test cases by default.
It would not make sense to keep the fuzzer running because if every test case crashes, all test cases need to be analyzed manually, which is unpractical and defeats the purpose of fuzzing.
The crashes are rooted in the fact that the fuzzer does not expect any value returned to the register, but in the case of register 23, it returns the written value.
Fortunately, this provides the strongest evidence for the presence of that specific register.
An additional indication for the existence of a register is if other registers -- which are always probed -- change.
Hence, for the unknown register 23, it is necessary to define crash settings to expect the written value in the register.
This starts the process of investigating the crashes, i.e., iteratively refining the crash settings and continuously learning more about the register and the functionality of the register bits.

\paragraph{Manual Analysis and Testing Hypothesis}
This iterative analysis includes inspecting the results and comparing the values written to the unknown register 23.
Also, we wrote a helper script that searches for identical bits in all transmitted values to identify bits that cause certain types of crashes.

Once we have observed specific bits influencing the state and formed a working hypothesis.
We then used the following method to test it.
The identified bit is fixated in the fuzzing mask such that only the remaining bits are fuzzed.
The crash settings are defined to reflect the presumed hypothesis.
Then, if no crash occurs, the working assumption has been confirmed: that the fixed bit has the expected influence on the \ac{FPGA} state.

Also, we can continue using \fwname as a rapid prototyping tool to manually investigate found crash cases by setting different static bits and changing the fuzzing masks accordingly.
Again, after gaining such information, it is possible to exclude a bit (or bits) from the default fuzzing process to continue fuzzing as initially.
Separately, the excluded bit or bit groups can be further investigated manually or with specific fuzzers.
Note that it is also possible to use the rapid prototyping feature of \fwname to encrypt the bitstream or insert additional commands, e.g., writes to the fabric, before or after the write to the investigated register, and observe if this changes the behavior of the target device.

\paragraph{Bit 24}
With this method, we first revealed that the \ac{FPGA} soft crashes if bit 24 is set, e.g., all registers return zero, but the \ac{FPGA} can be reset by the \texttt{JPROGRAM} \ac{JTAG} command.

\paragraph{Bit 16 and 17}
After this new information gain, we removed bit 24 from our fuzzing masks and investigated the remaining bits.
Analyzing the newly found crashes reveals an influence of bits 16 and 17 on the data read from the \texttt{FDRO} register, which is used to read back the fabric data, i.e., the \ac{FPGA} design.
As per the documentation, it is advised to read an additional frame when accessing the \texttt{FDRO} register, along with 10 words (in UltraScale) or 25 words (in UltraScale+) to consider pipelining~\cite{UG570}.
If bit 16 is set, single bits are set in the pipeline frame, and if bit 17 is set, the sync word can be seen in the pipeline words.
Unfortunately, we cannot comprehensively explain the results, which we will address in future work.

\paragraph{Bit 23, 25, and the Influence of a Programmed BBRAM Key}
We noticed by coincidence that some of the tested values in register 23 flag the \ac{BBRAM} key for deletion after a power cycle.
In other words, the BBRAM is cleared after testing the unknown register 23 and a power cycle.
Running the fuzzer again if \textit{no} \ac{BBRAM} key is programmed reveals that the \ac{FPGA} hard crashes if bit 23 and bit 25 are set.
Hence, no communication is possible with the \ac{FPGA}, i.e., it is unresponsive.
Only manually power cycling the device resets the \ac{FPGA}, an error state we only encountered during the investigation of the unknown register 23.
Further investigating this hard crash reveals that before the \ac{FPGA} goes into the unresponsive state, it returns 156 bytes of seemingly random data.
Re-executing this crashed test case and comparing the data shows similarities in that data.
For example, bits at certain positions are always identical.
We assume the returned data are kind of a crash dump in response to the value written to the unknown register 23 before the device locks in an error state.\\
For a more extensive list of our fuzzing results related to the unknown register 23 (and other registers), we refer to our GitHub repository~\cite{RapidBitFuzzGithub}.

\section{Discussion \& Future Work}\label{js::chapter::discussion}

The unpatchable nature of \ac{FPGA} configuration engines underscores the critical importance of proactive security measures during hardware systems design and development. 
In this work, we highlight the effectiveness and efficiency of hardware fuzzing by automated identification of critical vulnerabilities among other safety and reliability-threatening states, as well as contributing to the general understanding of the configuration engine, thus addressing our research question.

\paragraph{Security Implications}
With the unpatchable JustSTART attack, we circumvent the \ac{RSA} bitstream authentication.
This vulnerability enables attackers to load non-encrypted bitstreams containing hardware Trojans or manipulate a plaintext \ac{RSA}-authenticated bitstream, resulting in the leakage of sensitive data during runtime.
We want to highlight that this attack can be mitigated by enforcing \ac{AES} encryption using security fuses, as our attack does not bypass the decryption process.
So, JustSTART only breaks authentication, but it does not break confidentiality. 
In contrast, the starbleed attack targets bitstream confidentiality and is mitigated in case the bitstream is authenticated. Thus, mitigating both attacks simultaneously is only possible by enforcing both \textbf{authentication and confidentiality} using the security fuses, as already highlighted in~\cite{EnderLMP22}.

\paragraph{FPGA Bitstream Authentication vs. Encryption}
\ac{FPGA} bitstream protection schemes typically offer both authenticity and confidentiality as security properties.
In various scenarios, authentication-only is a valid security goal as \ac{FPGA} design confidentiality is not required, e.g., open-source \ac{FPGA} designs where the \ac{HDL} is openly available anyways, or when confidentiality is not a functional (user) requirement.
For instance, the patchable bitstream encryption engine~\cite{Unterstein:2019:SSU:3338508.3359573} requires only authentication for not being manipulated.
As another example, the recent DoD report~\cite{DoDFPGAAssurance} thoroughly analyzes potential threats and countermeasures for \ac{FPGA} assurance.
For instance, to prevent adversaries from swapping the bitstream (cf. TD~6), they only recommend authentication.
Generally, the report specifically excludes scenarios where confidentiality is required, which are addressed in other reports.
We also want to highlight that \ac{FPGA} security configuration errors by hardware designers may lead to authentication-only bitstream protection due to dispersed and inconsistent official documentation, cf.~\cite{ICAPPaper}.
Xilinx does not clearly recommend using authentication \emph{and} encryption, i.e., XAPP1098~\cite{XAPP1098} discusses the use of \textit{authentication of unencrypted bitstreams}.
Based on our research, Xilinx will update its public documentation as part of the vulnerability disclosure process.

\paragraph{\fwname Framework}

Our primary motivation for \fwname was providing a fuzzing framework with rapid prototyping capability to evaluate -- security-related -- configuration engine hypotheses with an automated workflow.
Note that Ender et al.~\cite{DBLP:conf/uss/Ender0P20, EnderLMP22} identified the starbleed attack by investigating the documentation and manual bitstream generation. 
The results discussed in \autoref{js::chapter::case-studies} show that we automatically rediscover starbleed (even working with other registers than the original attack), and we identified the novel JustSTART attack, resulting in a loss of authentication.
Moreover, we want to emphasize that besides the identification of vulnerabilities, a key motivation is to improve the understanding of the configuration engine since the official documentation is incomplete and dispersed.
We refer interested readers to~\cite{RapidBitFuzzGithub} for comprehensive evaluation results.

A key advantage of \fwname is that defining bitstreams in code is considerably less error-prone than manual construction using a hex editor. Note that this drastically reduces required time to construct mutated bitstreams and accelerates hypothesis evaluation.

For example, we leveraged this rapid prototype approach to quickly identify an undocumented \ac{RSA} test mode mentioned in a Xilinx patent~\cite{xilinx_rsa_patent}.
We also want to highlight that the underlying boofuzz framework~\cite{pereyda_boofuzz_2022} provides valuable features and enables fast implementation of necessary extensions to communicate with target devices and craft custom-tailored optimizations.

\paragraph{Systematic Fuzzing Strategies}
Our investigation into the \ac{FPGA} configuration engine was guided by a systematic and iterative approach (\autoref{js::chapter::strategies}), i.e., i) select and ii)~implement a fuzzer with the help of our three main strategies, iii) gather, and iv) evaluate \ac{FPGA} state information to then adjust fuzzer mutation accordingly.
Throughout this approach, we were able to establish an improved understanding of the internal workings of the configuration engine, uncovering vulnerabilities and shedding light on various aspects of its functionality.
However, we also want to highlight the limitations of such an iterative approach.\\
\begin{enumerate}[label=\roman*)]
    \item The approach depends on natural language interpretation of public documentation and patents, often lacking essential technical details such as clear concepts and implementation details.
This highlights the critical need for comprehensive documentation in facilitating effective security analysis.
    \item Since this process typically requires guidance by a security engineer, complex fuzzing sequences are not created automatically so far. As of now, this limits our evaluation and mostly sheds light on single registers rather than complex combinations and interactions of them (c.f. next paragraph).
    \item Most of our evaluation assists in understanding the configuration engine's inner workings to a limited extent, i.e., uncovering the behavior vs. purpose.
The fuzzing reveals certain inner working effects (i.e., the behavior of the configuration engine).
However, it does not help to explain them due to the limited internal status information (i.e., what is the purpose of a particular bit-field?).
Therefore, we believe running the current setup for a prolonged period may yield new crashes, but the absence of a comprehensive way to interpret results limits potential for novel insights.
\end{enumerate}

In future work, we plan to address these issues with feedback-based and coverage fuzzing strategies by incorporating \ac{FPGA} state information as feedback into mutation generation. Such a fuzzer should ideally automatically create new mutations based on feedback and coverage of bit patterns within the bitstream structure. During our evaluation, we observed that adjusting crash settings rather than using a new bitstream mutation strategy yielded more \textit{useful} crash results.
Note that this may be due to the fact that we have focused on single configuration register observations so far rather than complex register state interactions.
To also address the third limitation, future work should focus on automatic generation of a (detailed) configuration engine state-machine model to enable an improved understanding of its inner workings.

\paragraph{Vulnerability Discovery}
We now highlight our vulnerability discovery process using our framework \fwname. Based on rapid prototyping design strategies, we built automated fuzzers to i) run randomized structure-aware fuzzers and ii) validate attack hypotheses by definition of explicit crash settings.\\
In case of (structure-aware) intra-command and structural fuzzers, we formulated the crash settings as deviating from default initialized values and then let the fuzzer run for a certain time until crashes occurred.
In case of the JustSTART vulnerability, we formulated the attack goal that the \ac{FPGA} boots an authenticated but manipulated bitstream.
We then defined this objective as a crash setting (cf. \autoref{js::lst::juststart_fuzzer}) in our fuzzing context and then fuzzed the configuration engine until a crash, i.e., an unauthorized boot-up occurred.
Note that in each crash report, \fwname automatically includes the corresponding bitstream responsible for said crash.
Similarly, we discussed extending JustSTART with (un-)setting the \texttt{DEC} bit in \autoref{js::chapter::case-studies::juststart::extending_attack}, where we narrowed the crash setting only to catch a set or reset of said bit.

From a high-level perspective, we observed that random fuzzing might yield vulnerabilities such as denial of service.
However, concrete crash formulation based on a guided idea yielded more effective attacks, as demonstrated in this work.

\paragraph{On HDL Fuzzing}
While \ac{HDL} fuzzing may appear more effective and efficient at first glance, confirmation of vulnerabilities (and their absence) post-silicon remains essential even if \ac{HDL} source code is at hand, particularly considering the difference between high-level \ac{HDL} and an optimized fabricated chip.
To this end, our framework \fwname can be integrated into a typical CI/CD pipeline.
Further, there are gray-box settings -- as our current one -- where only documentation and a chip without the \ac{HDL} code are available.

\paragraph{Performance}
Another limitation is the test-case performance rooted in the connection to devices under test.
The utilized \ac{JTAG} interface uses a single-bit interface, clocked at 66 MHz only.
In our experiments, we could test up to $2^{21}$ bitstreams per day on UltraScale(+) and on our 7-Series cluster $2^{25}$.
Reasonable enough to run fuzzers discovering undefined behavior and said attacks.
However, we are practically limited to fuzzing a complete 32-bit register or sequences of several commands consisting of fewer bits.
Note that the test-case performance depends on fuzzer configurations like bitstream length and crash settings.

For future research, one may evaluate an UltraScale(+) \ac{FPGA} device cluster or use the SelectMAP interface that employs a 32-bit bus and higher speed, theoretically being up to 66 times faster than \ac{JTAG}.
Furthermore, our work can be extended to analyze the \ac{JTAG} controller itself, as it is used for programming the \ac{BBRAM} key and security fuses.

\paragraph{Vendors}
We want to emphasize that our work is centered around Xilinx \acp{FPGA}, which share a basic architecture, particularly the configuration process and configuration packets.
Using \fwname, we demonstrated the possibility of rapidly generating and fuzzing bitstreams to discover vulnerabilities. 
We are confident this concept can be transferred to \acp{FPGA} from other vendors, so we plan to expand our work in this regard in the future.
However, keep in mind that processor-based systems such as Xilinx Zynq devices have to be handled with a different approach, i.e., processor-based fuzzer approaches such as sandsifter~\cite{sandsifter}.

\section{Conclusions}

In this work, we demonstrated the effectiveness and efficacy of hardware fuzzing for the Xilinx 7-Series and UltraScale(+) configuration engine with only the chip and limited documentation at hand.
We designed and implemented \fwname, a framework designed to facilitate the creation of structure-aware mutational fuzzers. 
Additionally, its rapid prototyping capabilities empower analysts to delve deeper into the fuzzing results and swiftly validate assumptions.
We formulated three primary fuzzing strategies to establish a systematic approach to fuzz Xilinx \ac{FPGA} configuration engines.
Besides rediscovering the starbleed~\cite{DBLP:conf/uss/Ender0P20, EnderLMP22} attack, we identified a new unpatchable vulnerability named \emph{JustSTART} (CVE-2023-20570) that bypasses \ac{RSA} authentication of all Xilinx UltraScale(+) \acp{FPGA}.
Moreover, we revealed various undocumented and unexpected behavior, e.g., an \ac{RSA} test mode or a bit pattern crashing the \acp{FPGA} into an unresponsive state.

Since we believe that our work raises awareness for hardware security designers and analysts, we released our fuzzing and rapid prototyping framework \fwname under the MIT license~\cite{RapidBitFuzzGithub}.

\section*{Acknowledgments}
This work was partly supported by the Deutsche Forschungsgemeinschaft (DFG, German Research Foundation) under Germany's Excellence Strategy (EXC 2092 CASA 390781972).
We also thank our reviewers for their helpful feedback and Xilinx for their excellent responsible disclosure process, which started on February 27th, 2023, and was acknowledged a day later.


\appendix

\section{Fuzzer Listings}

\begin{lstlisting}[language={python}, caption={Excerpt of the source code of the starbleed fuzzer.}, numbers=left, label=js::lst::starbleedrequest, float=htb]
custom_register_settings = {
    "DEFAULT": {
        "crash_if_differs_from_default": "no",
        "crash_if_equal_to": "F0 0D F0 0D, BE EF BE EF, DE AD C0 DE",
    },
    "register0": {
        # Overwrite the default crash setting from the default_register_settings.ini.
        "crash_if_equal_to": "F0 0D F0 0D, BE EF BE EF, DE AD C0 DE",
    },
    "register3": {
        # The FDRO register should only return zeros because encryption is enabled.
        "crash_if_differs_from_default": "yes",
        "crash_if_equal_to": "",
    },
    "register5": {
        # Overwrite the default crash setting from the default_register_settings.ini.
        "crash_if_not_equal_to": "",
    },
}

starbleed_request = Request(
    name="starbleed_request",
    children=(
        FuzzedBitstream(
            name="starbleed_bitstream",
            file_name="write_fdri_bbram_test_key.bit",
            fuzzing_mask=0xF803E7FF,
            fuzzing_position=FuzzPosition(index_start=284, word_count=40),
        )
    ),
)
\end{lstlisting}

\begin{lstlisting}[language={python},caption={Excerpt of the source code of the JustSTART fuzzer.}, numbers=left, label=js::lst::juststart_fuzzer]
custom_register_settings = {
    "DEFAULT": {"probe": "no"},
    "register7": {
        "probe": "yes",
        "crash_if_differs_from_default": "no",
        # Only crash if BIT13_DONE_INTERNAL_SIGNAL_STATUS or BIT14_DONE_PIN is set.
        "crash_if_some_bits_in_mask_set": "00 00 C0 00",
    },
    "register22": {
        "probe": "yes",
        "crash_if_differs_from_default": "no",
        # Only crash if just BIT00_STATUS_VALID_0 is set.
        "crash_if_equal_to": "00 00 00 01",
    },
}

plaintext_rsa_bitstream_request = Request(
    name="plaintext_rsa_bitstream_request",
    children=(
        # Disable ConfigFallback in the CTL0 register.
        Type1WritePacket(name="write_to_mask", register_address=6),
        Static(name="mask_value", default_value=b"\x00\x00\x05\x01"),
        Type1WritePacket(name="write_to_ctl0", register_address=5),
        Static(name="ctl0_value", default_value=b"\x00\x00\x05\x01"),
        NOP(3),
        PlaintextRSABlockUltraScale(
            name="plaintext_rsa_block",
            children=(
                # Original RSA header, except ConfigFallback is disabled in the CTL0 register.
                NOP(),
                Type1WritePacket(name="write_to_mask_1", register_address=6),
                Static(name="mask_value_1", default_value=b"\xFF\xFF\xFF\xFF"),
                Type1WritePacket(name="write_to_ctl0_1", register_address=5),
                Static(name="ctl0_value_1", default_value=b"\x00\x00\x05\x01"),
                Type1WritePacket(name="write_to_mask_2", register_address=6),
                Static(name="mask_value_2", default_value=b"\xFF\xF3\xFF\xFF"),
                Type1WritePacket(name="write_to_ctl1", register_address=24),
                Static(name="ctl1_value", default_value=b"\x00\x00\x00\x00"),
                NOP(8),
                Type1WritePacket(name="write_to_far_1", register_address=1),
                Static(name="far_value_1", default_value=b"\x00\x00\x00\x00"),
                Type1WritePacket(name="write_to_cmd_1", register_address=4),
                Static(name="wcfg_code", default_value=b"\x00\x00\x00\x01"),
                NOP(11),
                # 25 or 26 frames of fabric data for plaintext test mode RSA bitstreams.
                Static(
                    name="fabric_data",
                    default_value=b"\xDE\xAD\xC0\xDE" * CONSTANTS.BOARD_CONSTANTS.FRAME_LENGTH * 25,
                ),
                # Original RSA footer, except ConfigFallback is disabled in the CTL0 register.
                NOP(2),
                Type1WritePacket(name="write_to_cmd_2", register_address=4),
                Static(name="grestore_code", default_value=b"\x00\x00\x00\x0A"),
                NOP(2),
                Type1WritePacket(name="write_to_cmd_3", register_address=4),
                Static(name="dghigh_code", default_value=b"\x00\x00\x00\x03"),
                NOP(20),
                Type1WritePacket(name="write_to_cmd_4", register_address=4),
                Static(name="start_code", default_value=b"\x00\x00\x00\x05"),
                NOP(),
                Type1WritePacket(name="write_to_far_2", register_address=1),
                Static(name="far_value_2", default_value=b"\x07\xFC\x00\x00"),
                Type1WritePacket(name="write_to_mask_3", register_address=6),
                Static(name="mask_value_3", default_value=b"\x00\x00\x05\x01"),
                Type1WritePacket(name="write_to_ctl0_2", register_address=5),
                Static(name="ctl0_value_2", default_value=b"\x00\x00\x05\x01"),
                NOP(2),
                Type1WritePacket(name="write_to_cmd_5", register_address=4),
                Static(name="desync_code", default_value=b"\x00\x00\x00\x0D"),
                NOP(119),
            ),
            children_contain_header_and_footer=True,
            key_file_name="test_key_rsa.nky",
            rsa_private_key_file_name="privateKey_wrong.pem",
            test_mode=True,
            rdw_go=False,
        ),
        Type1WritePacket(name="write_to_cmd_1", register_address=4),
            BitstreamWord(
                name="fuzzed_cmd_value_1", static_bits=0x00000000, fuzzing_mask=0x0000001F,
            ),
            NOP(3),
            Type1WritePacket(name="write_to_cmd_2", register_address=4),
            BitstreamWord(
                name="fuzzed_cmd_value_2", static_bits=0x00000000, fuzzing_mask=0x0000001F,
            ),
            NOP(3),
            Type1WritePacket(name="write_to_cmd_3", register_address=4),
            BitstreamWord(
                name="fuzzed_cmd_value_3", static_bits=0x00000000, fuzzing_mask=0x0000001F,
            ),
        NOP(3),
        Type1WritePacket(name="write_to_cmd", register_address=4),
        Static(name="rdw_go_code", default_value=b"\x00\x00\x00\x16"),
        NOP(3),
    ),
)
\end{lstlisting}

\bibliographystyle{alpha}
\bibliography{refs}

\end{document}